\documentstyle[12pt,epsf]{article}

\newcommand{\beq}{\begin{equation}}
\newcommand{\eeq}{\end{equation}}
\newcommand{\nn}{\nonumber}
\newcommand{\bea}{\begin{eqnarray}}
\newcommand{\eea}{\end{eqnarray}}

\newcommand{\gtrsim}{\ \rlap{\raise 2pt\hbox{$>$}}{\lower 2pt \hbox{$\sim$}}\ }
\newcommand{\lessim}{\ \rlap{\raise 2pt\hbox{$<$}}{\lower 2pt \hbox{$\sim$}}\ }

\newcommand{\np}[1]{Nucl. Phys. {\bf #1}}
\newcommand{\pl}[1]{Phys. Lett. {\bf #1}}
\newcommand{\pr}[1]{Phys. Rev. {\bf #1}}

\newcommand{\ijmp}[1]{Int. Jour. Mod. Phys. {\bf #1}}
\newcommand{\mpl}[1]{Mod. Phys. Lett. {\bf #1}} 
\newcommand{\ptp}[1]{Prog. Theor. Phys. {\bf #1}}

\makeatletter
\if@twoside
\else \oddsidemargin 00pt \evensidemargin 00pt
\fi
\let\@eqnsel = \hfil

\expandafter\ifx\csname mathrm\endcsname\relax\def\mathrm#1{{\rm #1}}\fi
\@ifundefined{mathrm}{\def\mathrm#1{{\rm #1}}}{\relax}

\makeatother

\begin{document}
\thispagestyle{empty}
\null
\hfill FTUV/97-53,IFIC/97-54

\hfill hep-ph/9708209

\vskip 1.5cm

\begin{center}
{\Large \bf      
$\Delta F=2$ Effective Lagrangian in theories with vector-like fermions
\par} \vskip 2.em
{\large         
{\sc G. Barenboim and F.J.Botella
}  \\[1ex] 
{Departament de F\'\i sica Te\`orica, Universitat 
de Val\`encia  and  IFIC, Centre Mixte Universitat 
de Val\`encia - CSIC} \\
{\it E-46100 Burjassot, Valencia, Spain} \\[1ex]
\vskip 0.5em
\par} 
\end{center} \par
\vfil
{\bf Abstract} \par
In this work we analyze a new piece 
present in the $\Delta F = 2$ effective Lagrangian
in models with extra vector-like quarks.
This piece, which was not taken into account previously, 
is required in order to preserve gauge invariance once
the unitarity of the CKM matrix is lost.
We illustrate the effects of this new piece in both, CP conserving and CP
violating processes.
\par
\vskip 0.5cm
\noindent July 1997 \par
\null
\setcounter{page}{0}
\clearpage

\section{Introduction}

In the Standard Theory of the Electroweak Interactions(ST) there remain two
main topics to be understood: the spontaneous breaking of the gauge symmetry
(the Higgs sector) and the origin of the families (the flavour problem,
including CP violation).The first problem will be studied with great detail
once the hadron machines in the TeV region (LHC) start to operate.
Nevertheless to get more insight in the flavour problem, it is not now
mandatory to explore new energy regions. Instead, by reaching high
luminosity in ``low'' energy machines such as tau-charm or beauty factories,
one can perform tests of the ST in the flavour sector, not previously done,
and consequently try to look for potential new physics.

There are several theoretically well-founded extensions of the Standard
Theory, not excluded by experimental data, and giving rise to large
deviations of the ST predictions in the flavour sector. In this paper we are
interested in models with an extended quark sector. In particular, models
with additional vector-like quarks 
have been extensively studied in the
literature \cite{vlq2,gustavo}. 
The most salient feature of this kind of theories is the
presence of $Z^0$-mediated flavour changing neutral currents(FCNC) at tree
level. The origin of FCNC can be traced back to the failure of unitarity of
the Cabbibo-Kobayashi-Maskawa(CKM) mixing matrix V. For simplicity, if we
add to the ST just a vector-like singlet quark with charge $Q=-\frac 13$,
the down quark mass eigenstates will be a mixture of the four down quarks
present in the model and consequently V will be a 3 x 4 submatrix of a 4 x 4
unitary matrix. The columns of V are no more ortogonal, therefore the model
has $Z^0$-mediated FCNC in the down sector.

The usual strategy to confront these kind of models with experiments has
been to take as effective Lagrangian for $\Delta F=1,2$ FCNC processes, the
ST Lagrangian (one loop) plus the new tree-level contribution. Of course,
when the experimental data are analyzed with this new Lagrangian,
significant deviations of ST values of the matrix elements $V_{ij}$ can be
obtained, and this fact has been taken into account as is the case for $%
V_{td}$ in the analysis of $B_d^0-\overline{B_d^0}$ mixing. What never has
been taken into account in the {\em calculation} of the effective
Lagrangian, is the deviations from unitarity of the CKM-matrix, for example:

\begin{equation}
\sum_{i=1}^3V_{ib}^{*}V_{id}=U_{bd}\neq 0  \label{unit bd}
\end{equation}
if $U_{bd}$ is different from zero, the box diagram for $B_d^0-\overline{%
B_d^0}$ {\em is not gauge invariant}, implying the existence of new
contributions, previously not considered. Even worse, the $Z^0$-penguin is 
{\em divergent}, so a reanalysis of $\Delta F=2$ boxes and $\Delta F=1$
electroweak penguins is mandatory in this class of theories. In fact is
quite easy to understand that the new pieces in the $\Delta F=2$ can be {\em %
linear} in the new physics($U_{bd}$), contrary to the tree level
contribution that is {\em quadratic} in $U_{bd}$. In the ST $%
V_{ub}V_{ud}^{*} $ is changed by minus the sum of the c and t quark
couplings, giving rise to cancellations of gauge dependent pieces( at the
same time that the ultraviolet behavior of the graph is improved). In these
theories and following Eq.(\ref{unit bd}) there remains a gauge-dependent
piece proportional to $U_{bd}$, coming from the box. Of course new pieces
must be present to make gauge invariant this linear piece in $U_{bd}$, but
in any case it is evident that in the limit of small $U_{bd}$, a linear term
will dominate over the quadratic one . So we can expect that this new
contributions will become important when both ST and new physics give rise
to contributions of the same order.

The goal of this paper is to find the pieces linear in $U_{bd}$, present in
the $\Delta F=2$ effective Lagrangian, in theories with extra vector-like
singlet quarks. In section 2 we present a brief review of the model in
order to fix the notation. Section 3 is devoted to present the calculation
of the new pieces in the $\Delta F=2$ effective Lagrangian, also some
comments about the $\Delta F=1$ pieces are included. In section 4 we
illustrate the numerical effect of the presence of the new piece linear in $%
U_{qp}$ both for CP conserving and CP violating processes. And finally in
section 5 we present our conclusions, where it is stressed the necessity of
making a general numerical reanalysis of the predictions of these kind of
models.

\section{Models with vector-like quarks}

For simplicity we will take the standard $SU(2)\otimes U(1)$ gauge theory
with the addition of one $Q=-\frac{1}{3}$ down vector-like quark, singlet
under $SU(2)$. Therefore the quark content of the model will be in the weak
basis: 
\begin{equation}
\begin{array}{cccc}
q_{Li}^{0}\equiv \left( 
\begin{array}{c}
u_{L}^{0} \\ 
d_{L}^{0}
\end{array}
\right) _{i} & u_{Ri}^{0} & d_{R\alpha }^{0} & d_{L4}^{0} \\ 
 \\
(1/2,1/3) & (0,4/3) & (0,-2/3) & (0,-2/3)
\end{array}
\label{quark content}
\end{equation}
where i=1,2,3 and $\alpha $=1,2,3,4, and the weak isospin and hypercharges
have been written explicitly. The Yukawa sector of the model with the
standard Higgs-doublet $H$ is: 
\begin{equation}
{\cal L}_{Y}=\overline{q}_{Li}^{0}\Gamma _{i\alpha }d_{R\alpha }^{0}H+%
\overline{q}_{Li}^{0}\Delta _{ij}u_{Rj}^{0}\widetilde{H}+\overline{d}%
_{L4}^{0}\Sigma _{\alpha }d_{R\alpha }^{0}+h.c.  \label{Yukawa Lagra}
\end{equation}
in general one can take $\Delta _{ij}$ real and diagonal in such a way that
the up weak basis is the same than the up mass basis. For the down sector we
have to diagonalize the mass matrix$\left( 
\begin{array}{c}
\Gamma \\ 
\Sigma
\end{array}
\right) $, what can be done through 
\begin{equation}
\begin{array}{ccc}
d_{L}^{0}=Ad_{L} & ; & A=\left( 
\begin{array}{c}
V \\ 
B
\end{array}
\right)
\end{array}
\label{down diagon}
\end{equation}
and an irrelevant rotation of $d_{R}^{0}$ .$A$ is a (4x4) unitary matrix and 
$V$ is a (3x4) submatrix that will play the role of the CKM-matrix. The weak
gauge currents can be written as : 
\begin{eqnarray}
{\cal L}_{W} &=&\frac{g}{\sqrt{2}}\overline{u}_{Li}V_{i\alpha }\gamma ^{\mu
}d_{L\alpha }W_{\mu }  \nonumber \\
{\cal L}_{Z} &=&\frac{g}{2\cos \theta _{W}}[\overline{u}_{Li}\gamma ^{\mu
}u_{Li}-\overline{d}_{L\alpha }U_{\alpha \beta }\gamma ^{\mu }d_{L\beta } 
\nonumber \\
&&-2\sin ^{2}\theta _{W}J_{em}^{\mu }]Z_{\mu }  \label{gauge Lagra}
\end{eqnarray}
Now, the charged currents mixing matrix $V$ is a (3x4) CKM-matrix and the
neutral current Lagrangian is not flavour conserving in the left-handed down
sector owing to the fact that $d_{L}$ is vector-like. In this case 
\begin{equation}
U_{\alpha \beta }=V_{\alpha i}^{\dagger }V_{i\beta }=\delta _{\alpha \beta
}-B_{\alpha }^{*}B_{\beta }  \label{unit alfabe}
\end{equation}
where unitarity of $A$ has been used. A similar result is obtained for the
Higgs couplings, 
\begin{equation}
{\cal L}_{H}=-\frac{g}{2M_{W}}\left[ \overline{u}_{Li}m_{i}^{u}u_{Ri}+%
\overline{d}_{L\alpha }U_{\alpha \beta }m_{\beta }^{d}d_{R\beta }\right]
H+h.c.  \label{Higgs Lagra}
\end{equation}
and again the coupling to the down quark does not conserve flavour, $%
m_{i}^{u} $ and $m_{\beta }^{d}$ are the up and down quark masses
respectively. In summary, the main differences with the ST are the existence
of four down quarks, the CKM-matrix is a (3x4) and consequently there
appears FCNC in the down sector, both in the gauge and the Higgs couplings.
As far as the independent flavour parameters are concerned, we can say that
the counting is six angles and three phases, like in a standard
four-generations model. Consequently one can use a standard-like
parametrization of the (4x4) A matrix \cite{quico}
or other one that has been advocated
in the literature \cite{paco}.

\section{$\Delta F=2$ Lagrangian}

From Eq.(\ref{gauge Lagra}) it is evident that the graphs of figure 1 will
contribute, at tree level, to all the processes where the flavour of the down
quark $\alpha $ changes in two units (and so does $\beta $ too,$\alpha \neq
\beta $). These graphs give rise to an effective $\Delta F=2$ Lagrangian of
the form 
\begin{equation}
{\cal L}_{eff}^{Z}=-\frac{G_{F}}{\sqrt{2}}\left( \overline{\beta }_{L}\gamma
^{\mu }\alpha _{L}\right) \left( \overline{\beta }_{L}\gamma _{\mu }\alpha
_{L}\right) U_{\beta \alpha }^{2}  \label{tree Lagra}
\end{equation}
This piece is the new physics contribution that usually has been added to
the ST-term in order to confront this model to the experimental data. The so
called ST-contribution (of course at one loop) comes from the diagrams in
figure 2. If the calculation is performed in a general $R_{\xi }$-gauge
(including the graphs with the unphysical Higgses) the contribution we get
is of the form 
\begin{equation}
{\cal L}_{eff}^{box}=-\frac{G_{F}\alpha }{\sqrt{2}4\pi \sin ^{2}\theta _{W}}%
\left( \overline{\beta }_{L}\gamma ^{\mu }\alpha _{L}\right) \left( 
\overline{\beta }_{L}\gamma _{\mu }\alpha _{L}\right) F_{\alpha \beta }
\label{box Lagra}
\end{equation}
where $F_{\alpha \beta }$ is given by 
\begin{equation}
F_{\alpha \beta }=2\sum_{i,j=1}^{3}V_{i\beta }^{*}V_{i\alpha }V_{j\beta
}^{*}V_{j\beta }F(i,j,\xi )  \label{Falfabeta}
\end{equation}
and $F(i,j,\xi )$ in the zero external momenta and masses limit is a
function of the masses of the up-quarks of type $i$ and $j$, also a function
of $M_{W}$ and dependent on the gauge parameter $\xi $ and consequently {\em %
not gauge-invariant(!)}. Note that $V$ is not unitary in this model,
therefore Eq.(\ref{Falfabeta}) is different from the ST-result. In fact if we
make use of the definition of $U_{\alpha \beta }$ in Eq.(\ref{unit alfabe}),
it is straightforward to get 
\begin{eqnarray}
F_{\alpha \beta } &=&2 
\sum_{i,j=2}^{3}V_{i\beta }^{*}V_{i\alpha }V_{j\alpha
}^{*}V_{j\beta }[F(i,j,\xi )-F(i,1,\xi )-F(1,j,\xi )  \nonumber \\
&&+F(1,1,\xi )]+2U_{\beta \alpha }\sum_{i=2}^{3}V_{i\beta }^{*}V_{i\alpha
}\left[ F(i,1,\xi )-F(1,1,\xi )\right]  \nonumber \\
&&+2U_{\beta \alpha }\sum_{j=2}^{3}V_{j\beta }^{*}V_{j\alpha }\left[
F(1,j,\xi )-F(1,1,\xi )\right]  \nonumber \\
&&+2U_{\beta \alpha }^{2}F(1,1,\xi )  \label{laformula}
\end{eqnarray}

The first piece (the double summatory) is gauge invariant and is the only
piece that has been included in this kind of models. It is the Inami-Lim
ST-box contribution 
\cite{loop}, except for the values of the CKM-matrix to be used.
From the point of view of gauge invariance, it is quite legitimate to use
only this piece from Eq.(\ref{laformula}), but the important question that
immediately arises from Eq.(\ref{laformula}) is if the other pieces can be
bigger than the first one or even bigger than the tree level result. Before
answering these questions we must stress that the terms in Eq.(\ref{laformula}%
) linear and quadratic in $U_{\beta \alpha }$ are not gauge-invariant, so if
these pieces can be important we must look for a new set of diagrams to
restore the gauge invariance of these contributions.

Coming back to the previous question it is quite clear that in the limit in
which the forth down quark decouples, we recover the ST-model and
consequently $U_{\beta \alpha }$($\beta \neq \alpha $) goes to zero, but not 
$V_{i\beta }^{*}V_{i\alpha }$ for $\alpha ,\beta =1,2,3$.This means that the
pieces quadratic with $U_{\beta \alpha }$ goes to zero more rapidly than the
terms lineal in $U_{\beta \alpha }V_{i\beta }^{*}V_{i\alpha }$. Therefore,
the pieces lineal in $U_{\beta \alpha }$ in Eq.(\ref{laformula}) will
dominate the tree level Lagrangian in Eq.(\ref{tree Lagra}) at least in the
small $U_{\beta \alpha }$ regime. So, a priori, there is not reason (except
probably the loop expansion) to consider the tree level $\Delta F=2$
Lagrangian in Eq.(\ref{tree Lagra}) to be the leading new physics
contribution and not the linear pieces in $U_{\beta \alpha }$ appearing in
Eq.(\ref{box Lagra}). At the end it will be the experimental precision that
will dictate which piece is more important, but from the point of view of a
perturbative treatment, it looks evident that a piece linear in the new
physics must be more important than a quadratic one, provided we have enough
experimental precision.

As far as the $U_{\beta \alpha }^{2}$ piece in Eq.(\ref{laformula}) is
concerned, we must point out that its contribution to Eq.(\ref{box Lagra}) is
of the order $G_{F}\alpha U_{\beta \alpha }^{2}$ compared to the tree level
that is of order $G_{F}U_{\beta \alpha }^{2}$. Of course having FCNC at tree
level, at one loop level we get radiative corrections to the tree level
coupling, and this new piece is an order $\alpha $ correction to Eq.(\ref
{tree Lagra}). So we will neglect this piece, we are not interested in the
full one loop renormalization of this model, but in the leading corrections
to the ST-result in any regime of the parameters. Our next task is to find
out the other one loop contributions of order $U_{\beta \alpha }V_{i\beta
}^{*}V_{i\alpha }$ and such that when summed up to the second and third
pieces in Eq.(\ref{laformula}) we get a gauge-invariant result. These other
diagrams must have a $Z$-flavour changing vertex and another vertex where the
change of flavour is generated by two $W$-vertices, so the kind of graphs we
are looking for are those depicted in figure 3. The blob with a $W$ in
these graphs means to attach a $W$ in the blob in all possible ways to the
fermion line, in order to construct a one loop diagram. So the $W$ will give
us the $V_{i\beta }^{*}V_{i\alpha }$ term, the $Z$ in the other vertex will
introduce $U_{\beta \alpha }$. Therefore any introduction of new physics in
the blob will contribute to a subdominant piece and consequently the blob in
figure 3 must be calculated like a ST-diagram. Note that if the $Z$
couples in the blob to an external line, we have to take $U_{\beta \gamma
}=\delta _{\beta \gamma }$ ($\alpha =1,2,3$), otherwise we would get a $%
U_{\beta \alpha }U_{\beta \gamma }V_{i\gamma }^{*}V_{i\alpha }$ piece that
is subleading. Similar, in the blob we have to consider $\sum_{i=1}^{3}V_{i%
\beta }^{*}V_{i\alpha }=\delta _{\beta \alpha }$, the remaining piece $%
B_{\beta }^{*}B_{\alpha }=U_{\beta \alpha }$ ($\alpha \neq \beta $) would
give us $G_{F}\alpha U_{\beta \alpha }^{2}$ and therefore subleading again.

Now we have to show that the sum of the contribution of figure 3 and the
two linear terms in $U_{\beta \alpha }$ of Eq.(\ref{laformula}) gives a
gauge-invariant result. First of all, the four diagrams in figure 3 give
the same contribution, so we can concentrate in the first one. Second, both
graphs in figure 2 gives the same contribution, so we can eliminate the
global factor of 2 in Eq.(\ref{laformula}) and concentrate in the first graph
of this figure. In addition , the two terms linear in $U_{\beta \alpha }$ in
Eq.(\ref{laformula}) are equal because $F(i,j,\xi )$ is symmetric in $i,j$.
Therefore we have to prove that the piece $U_{\beta \alpha
}\sum_{i=2}^{3}V_{i\beta }^{*}V_{i\alpha }[F(i,1,\xi )-F(1,1,\xi )]$ and the
first graph of figure 3 sum up to a gauge-invariant result.

But this has been proved long time ago, again in the Inami-Lim`s paper
\cite{loop},
because the sum is exactly the same sum that has to be done to calculate the
short-distance effective Lagrangian for $K\rightarrow \mu \overline{\mu }$.
In fact the sum of this two graphs gives the same result as the
coefficient of the operator $(\overline{\beta}_{L}\gamma _{\mu }\alpha
_{L})(\overline{\mu }_{L}\gamma ^{\mu }\mu _{L})$ but multiplied by $%
U_{\beta \alpha }$. Note that the piece we are talking about from the first
graph in figure 2 correspond to perform the sum of $i$ using ``unitarity''
of $V$ and fixing $j=1$ ($m_{1}^{u}=0$ for the up-quark) and finally
eliminating the two CKM-matrix from the $\beta j$ and $\alpha j$ vertices
and multiply the global result by $U_{\beta \alpha }$. This box in the limit
of zero external masses and momenta is the same box as the one we get for $%
\alpha \overline{\beta }\rightarrow \mu \overline{\mu }$ except for the
global $U_{\beta \alpha }$ factor. The same happens for the $Z$-exchange
graph, except that one must be careful in taking from the $\alpha \overline{%
\beta }\rightarrow \mu \overline{\mu }$ amplitude only the piece that is
proportional to the weak isospin of the muon. Note that this piece is pure
left-handed, while the piece proportional to the charge is pure vector and
does not contribute to $K\rightarrow \mu \overline{\mu }$. Therefore we get
for the total piece linear in $U_{\beta \alpha }$ four times the piece that
Inami and Lim got for the coefficient of the operator $(\overline{s}%
_{L}\gamma ^{\mu }d_{L})(\overline{\mu }_{L}\gamma ^{\mu }\mu _{L})$, that
of course is gauge-invariant. But this is in the quark amplitude, at the
effective Lagrangian level we have to divide by 2x2.

Our full effective $\Delta F=2$ Lagrangian, neglecting external masses and
momenta will be: 
\begin{eqnarray}
{\cal L}_{eff}^{\Delta F=2} &=&\frac{G_{F}}{\sqrt{2}}\left( \overline{\beta }%
_{L}\gamma ^{\mu }\alpha _{L}\right) \left( \overline{\beta }_{L}
\gamma_{\mu }\alpha _{L}\right) \times  \nonumber \\
&&\times \left\{ -U_{\beta \alpha }^{2}+\frac{\alpha }{4\pi \sin ^{2}\theta
_{W}}\left[ 4\widetilde{C}U_{\beta \alpha }+\widetilde{E}\right] \right\}
\label{Result}
\end{eqnarray}
where $\widetilde{C}$ and $\widetilde{E}$ are defined in
reference \cite{loop} as 
\begin{eqnarray}
\widetilde{E} &=&\sum_{i,j=2}^{3}V_{i\beta }^{*}V_{i\alpha }V_{j\beta
}^{*}V_{j\alpha }\overline{E}(x_{i},x_{j})  \nonumber \\
\widetilde{C} &=&\sum_{i,j=2}^{3}V_{i\beta }^{*}V_{i\alpha }\overline{C}%
(x_{i},x_{1}=0)  \label{Limfunct}
\end{eqnarray}
and $x_{i}=\left( m_{i}^{u}/M_{W}\right) ^{2}$ as usual. Taking into account
the actual values of the quark masses, these functions can be approximated
by the following results 
\begin{eqnarray}
\overline{E}(x_{c},x_{c}) &\simeq &-x_{c}  \nonumber \\
\overline{E}(x_{t},x_{t}) &=&\frac{-4x_{t}+11x_{t}^{2}-x_{t}^{3}}{%
4(1-x_{t})^{2}}+\frac{3x_{t}^{3}\ln x_{t}}{2(1-x_{t})^{3}}  \nonumber \\
\overline{E}(x_{c},x_{t}) &=&\overline{E}(x_{t},x_{c})\simeq -x_{c}\left[
\ln \frac{x_{t}}{x_{c}}-\frac{3x_{t}}{4\left( 1-x_{t}\right) }-\frac{%
3x_{t}^{2}\ln x_{t}}{4(1-x_{t})^{2}}\right]  \nonumber \\
\overline{C}(x_{c}) &\simeq &x_{c}  \nonumber \\
\overline{C}(x_{t}) &=&\frac{x_{t}}{4}\left[ \frac{4-x_{t}}{1-x_{t}}+\frac{%
3x_{t}\ln x_{t}}{(1-x_{t})^{2}}\right]  \label{Buras}
\end{eqnarray}

Equation(\ref{Result}) is our main result. In particular it is the second
piece, linear in $U_{\beta \alpha }$, that we claim that has previously
neglected, and that a priori cannot be excluded in comparison to the first
one. In particular, in the small $U_{\beta \alpha }$ limit ( the small new
physics limit ), it is clear that this piece will be more important than the
tree level result.

At this point it must be also clear that the $Z$-mediated $\Delta F=1$
effective Lagrangian that usually has been used is perfectly correct because
in this case the new physics contribution is linear with $U_{\beta \alpha }$%
, and therefore any new physics introduced at one loop will be subleading.
This is not the case for gamma mediated FCNC. They do not exist at tree
level and consequently at one loop level one gets the ST-contribution and
additional leading new physics \cite{quico1}. 
This piece contributes for example to $%
b\rightarrow s\gamma $ and has been studied in \cite{quico2}.

\section{Numerical analysis}
In this section we will show the effects of our new contribution in Eq.(12).
It must be stressed that it is in the $B_d$ system where there is a 
bigger window for the type of models we are considering.
$B-\overline{B}$ mixing still can be dominated by the new 
contribution in the case 
of models with one extra down-type vector-like quark (dVLQ).
But as it was pointed out in the previous section, the new contribution 
has more chances to become important in the case that the new 
physics is relatively small.
It was shown in reference \cite{gustavo} that in CP asymmetries in $B$ decays
\cite{nir2}
there can be big effects of physics beyond the standard one.
The reason for that is that what is important in this case is the resulting 
phase in
$B^0_d-\overline{B}^0_d$ mixing.
With a 20\% of new physics in the mixing of $B_d$, the authors of reference
\cite{gustavo} showed spectacular effects for the CP asymmetry in the
$J/\psi \, K_S$ and $\pi\pi$ channels, because the phase of the
$U^2_{\beta\alpha}$ piece can be completly different to the
standard model phase. Now looking at Eq.(12), we see that we are
introducing a new piece $\tilde{C}U_{\beta\alpha}$ with a new phase.
If the balance of magnitudes is fortunate (or unfortunate)
enough, it could happen that we get an important enhancement of the new
physics, but in any case it seems quite clear that previous results
have to be revised. Therefore we will concentrate on the specific case
of CP asymmetries in $B_d$ decays.

Following reference \cite{gustavo} and using Eq.(12) we get
for the off diagonal term of the $B_d$ mixing matrix $M_{12}$
\bea
M_{12}=M_{12}^{(SM)} \Delta_{bd}^*
\eea
where $M_{12}^{(SM)}$ is the standard model contribution
and $\Delta_{bd}$ is given by
\bea
\Delta_{bd} = 1 + a r e^{i\phi} - b r^2 e^{2 i \phi}
\label{bemol}
\eea
where
\bea
r e^{i \phi} = \frac{U_{bd}}{\lambda_t} \;,
\eea
\bea
\lambda_t = V_{tb}^* V_{td} \; ,
\eea
\bea
a = \frac{4 \overline{C}(x_t)} {\overline{E}(x_t)} \; ,
\eea
and
\bea
b= \frac{ 4 \pi \sin^2\theta_W}{\alpha} \frac{1}{\overline{E}(x_t)}
\eea
for the actual top mass \cite{buras} we get 
\bea
a=-3.271 \;\; \mbox{and}\;\; b=-164.78 
\eea
If we take into account that the actual bound for $r$ is $
r\leq .3 $ (see reference \cite{nir}), we conclude that the 
contribution of figure 1 (the piece with $b$, quadratic in $r$) can be at
maximum 16 times bigger that the new piece (the piece with $a$,
linear in $r$).
But this is in the case where the new physics completely overwhelms
the standard model contribution. In the interesting case
studied by Branco et al. \cite{gustavo}, where $-br^2 =0.2$
($r\simeq0.035$) we get $ar=0.114$, therefore it is quite evident
that the new piece is going to be important in this regime.
Even more, if this 
20\% contributions give important effects, even in the case
where the tree graph is important, and the new one loop piece
is not too small we expect important deviations
 owing to the presence
of a new phase.

In this model we have for the CP asymmetry in the $J/\psi \,K_s$ and
$\pi\pi$ channels
\bea
a_J & \equiv &- \frac{\Gamma\left(B^0 \longrightarrow J/\psi \, K_s\right)
 - \Gamma\left(\overline{B}^0 \longrightarrow J/\psi \, K_s\right)}{
 \left(\sin(\Delta M t) \right) \left(
\Gamma\left(B^0 \longrightarrow J/\psi \, K_s\right) +
\Gamma\left(\overline{B}^0 \longrightarrow J/\psi \, K_s\right)
\right)}     \\ 
&&\nn \\
& = & \sin \left( 2\beta -\mbox{arg}(\Delta_{bd})\right) \nn
\eea
\bea
a_\pi & \equiv & -\frac{\Gamma\left(B^0 \longrightarrow \pi^+ \pi^-\right)
 - \Gamma\left(\overline{B}^0 \longrightarrow \pi^+ \pi^-\right)}{
 \left(\sin(\Delta M t ) \right) \left(
\Gamma\left(B^0 \longrightarrow \pi^+ \pi^-\right) +
\Gamma\left(\overline{B}^0 \longrightarrow \pi^+ \pi^-\right)
\right)}     \\
&&\nn \\
& = & \sin \left( 2\alpha +\mbox{arg}(\Delta_{bd})\right) \nn
\eea
where $\alpha$ and $\beta$ are defined as usual
\bea
\alpha \equiv \mbox{arg} \left(- \frac{V_{td} V_{tb}^*}{V_{ud} V_{ub}^*}
\right)
\eea
\bea
\beta \equiv 
\mbox{arg} \left( -  \frac{V_{cd} V_{cb}^*}{V_{td} V_{tb}^*}
\right) \;.
\eea
In order to plot $a_J$ and $a_\pi$ versus $\phi$ for a given value of
$r$, to see how the new physics can change the standard model
value and to clarify the effects of the new piece, let us explain
over the unitarity quadrangle, the experimental input we need.

From figure 4, it is clear that with $r$ and $\phi$, the shape
of the small upper traingle is fixed. 
We know experimentally two sides of the lower triangle $\mid
V_{cd} V_{cb}^* \mid$ and $\mid V_{ud} V_{ub}^* \mid$, therefore
to recontruct $\alpha$ and $\beta$ we need the third side
of the lower triangle. 
But this can be accomplished by fixing the size of the small
upper traingle through the experimental mixing parameter $x_d$
\bea
x_d =  \frac{G_F \alpha_{\mbox{{\small em}}}}{6 \pi \sqrt{2} \sin^2\theta_W} 
B_B f_B^2 m_B \eta_B \tau_B \mid 
\lambda_t \mid^2  \mid \overline{E}(x_t, x_t) \mid \mid \Delta_{bd} 
\mid
\label{xd}
\eea
For a given value of $r$ and $\phi$, this formula fixes $\mid 
\lambda_t \mid$, so we know the full quadrangle (if possible),
and therefore $\alpha$ and $\beta$.

In figures 5 and 6 we show the two asymmetries in a case similar
to the one represented in reference \cite{gustavo}. In dashed
lines we have plotted the actual prediction for the effective
Lagrangian used previously to this work. In solid line, we
present our new result. From the figures, it is quite evident
that the deviation respect to the old results can be bigger than
the expected sensitivities of B-factories, therefore we
conclude that the new piece must be taken into account
and therefore the whole Lagrangian in Eq.(12) must be
used to look for vector-like contributions in CP asymmetries
in $B$-decays.
For completeness we also present,in figure 7, the CP conserving quantity
$\mid \Delta_{bd} \mid$.

In figures 8 and 9 we represented a case where the new physics
is more important (bigger $r$, $r=0.1$) and our new contribution
should be less important from the moduli point of view.
But in this case we get even more spectacular results, as
previously announced because what is important in this case
is the presence of new phases. In the figures there are regions where
the unitarity quadrangle does not close and therefore there
are forbidden regions in the $\phi$ space.
Note that if we include the full Lagrangian (Eq.(12)) there
is a region around $\phi=\pi$ that now is striclty forbidden.
Near this region $a_\pi$ can change from -1 to 0 if we include our 
new piece.
                           
\section{Conclusions}
In this work we have analyzed the effects of the
previously omitted linear piece in $U_{bd}$ of 
the $\Delta F =2$ effective
Lagrangian in theories with extra vector-like singlet quarks.

This term arises when one takes into account
that due to the deviation of the CKM matrix from unitarity,
the box diagram is not gauge invariant  involving necessarily 
the presence of new contributions, not
considered previously. This new piece, which is linear in $r$,
has been shown to be important, not only in the small $r$
regime, but also in the big $r$ one.
The reason was already given, it is because the new piece 
carries its own phase, which can be (and in general is)
different from both the quadratic and standard model term phases.

Special emphasis was given to the consequences of this new piece
for CP asymmetries in $B^0$ decays, 
which certainly can be quite important.
In light of this, it seems clear that previous results have to
be revised.

\vspace{.5cm}

\begin{center}
{\bf ACKNOWLEDGMENTS}
\end{center}
 It is a pleasure to thank O. Vives for his cooperation
 in part of this work and  G. Branco for enlightening discussions. 
We would also like to thank J. Bernab\'eu and F. del Aguila
for useful comments and
discusions.
G.B. acknowledges the Spanish Ministry of
Foreign Affairs for a MUTIS fellowship .
 This work is supported by CICYT under 
grant AEN-96-1718 and IVEI under grant 038/96.

\clearpage

\begin{figure}
\begin{center}
\epsfxsize = 13cm
\epsffile{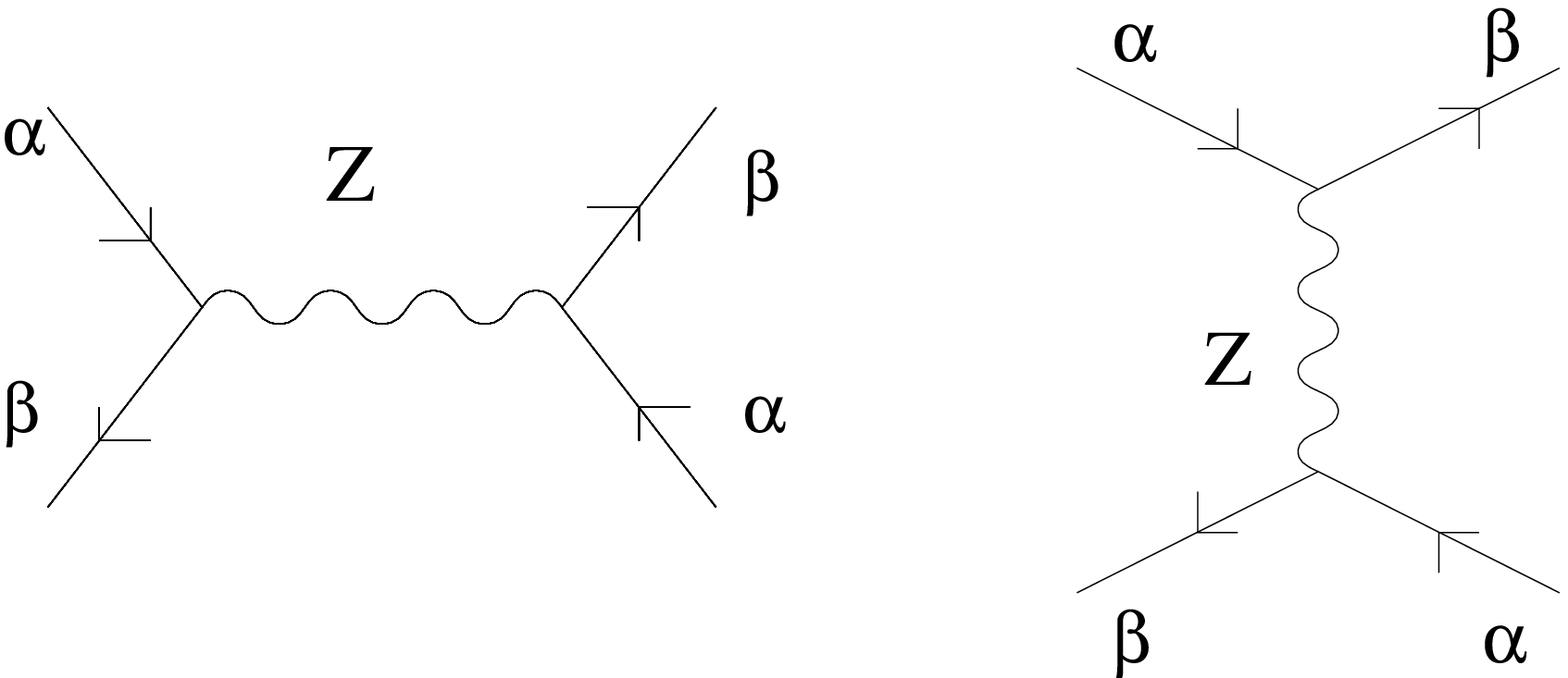}
\leavevmode
\end{center}
\caption{Tree level contributions  to the $\Delta F =2$ Lagragian. 
These pieces are the new physics contribution that 
usually have been taken
into account. }
\end{figure}

\clearpage

\begin{figure}
\begin{center}
\epsfxsize = 15cm
\epsffile{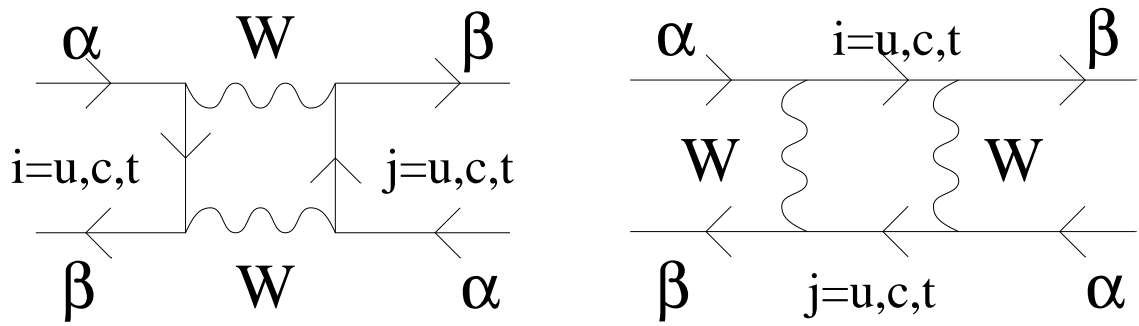}
\leavevmode
\end{center}
\caption{ Standard Model contribution to the $\Delta F = 2 $ effective
Lagrangian }
\end{figure}

\clearpage

\begin{figure}
\begin{center}
\epsfxsize = 13cm
\epsffile{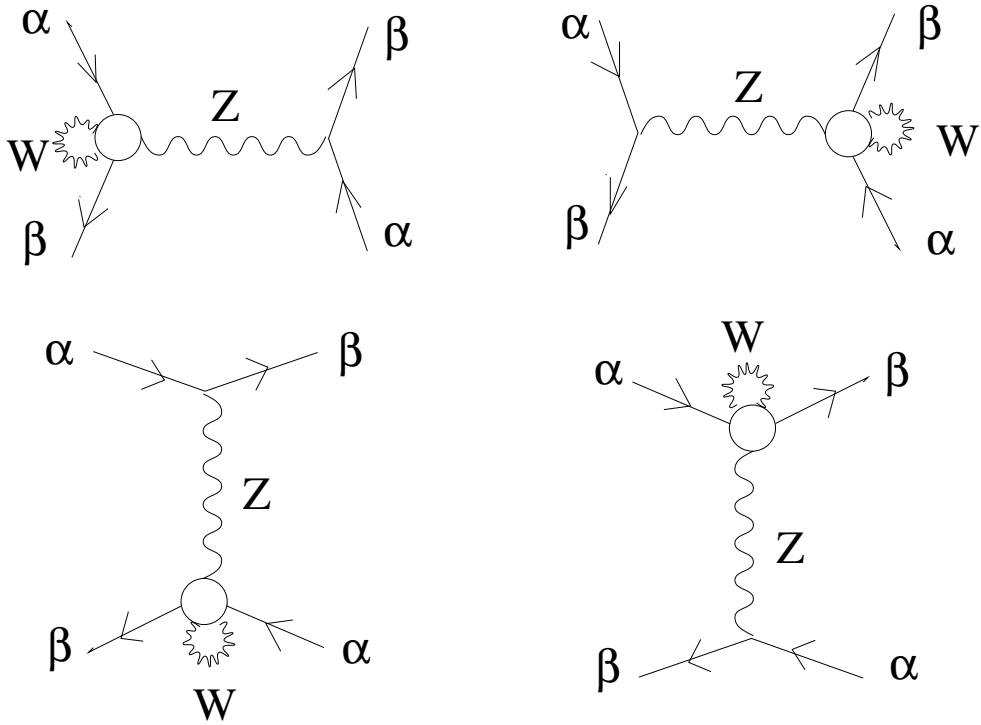}
\leavevmode
\end{center}
\caption{New pieces which are linear in $U_{\beta \alpha}$ }
\end{figure}

\pagebreak
\begin{figure}
\begin{center}
\epsfxsize = 13cm
\epsffile{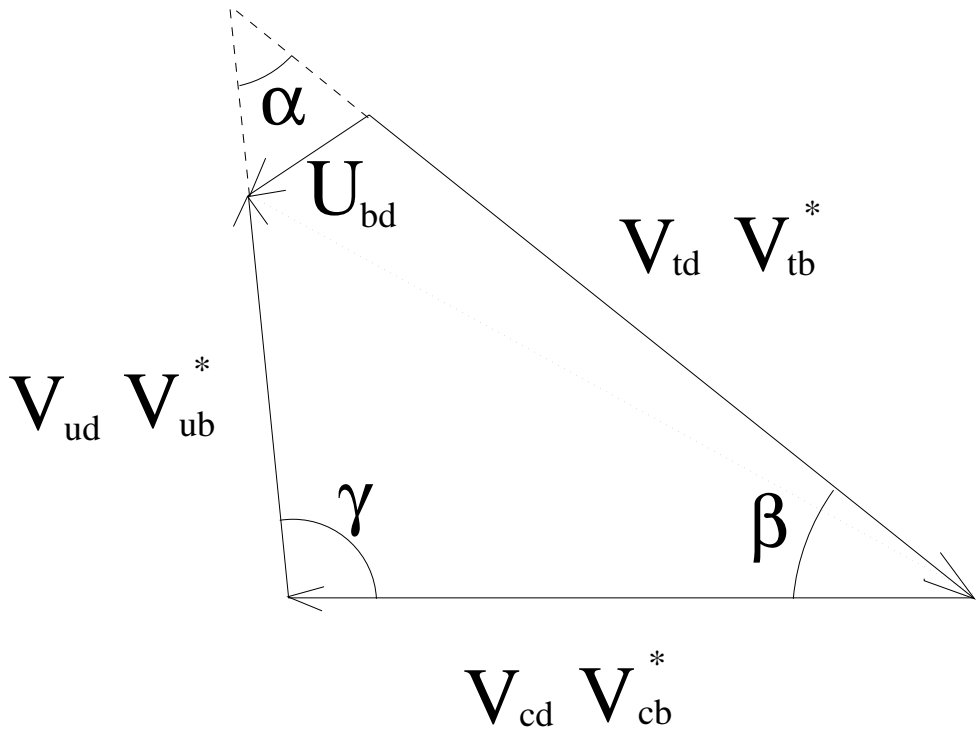}
\leavevmode
\end{center}
\caption{Unitarity quadrangle in the $B_d$ sector }
\end{figure}

\clearpage

\begin{figure}
\begin{center}
\epsfxsize = 13cm
\epsffile{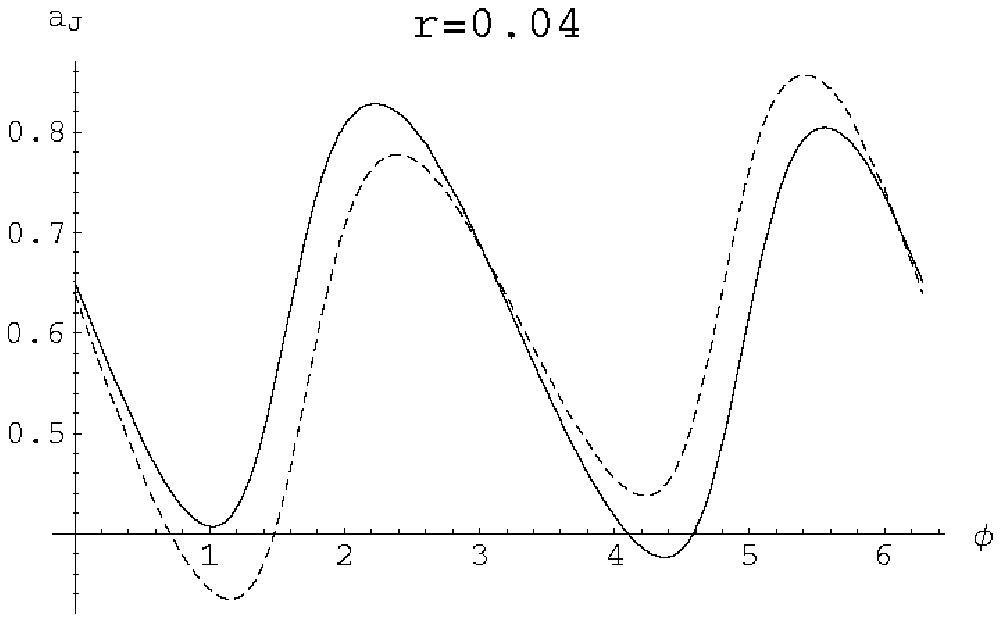}
\leavevmode
\end{center}
\caption{$a_J$ as a function of $\phi$ for $r=0.04$. 
The dashed line  corresponds to the prediction obtained with  
the effective  Lagrangian previously used 
(without the linear term), while the
solid line corresponds to our new result.}
\end{figure}

\clearpage

\begin{figure}
\begin{center}
\epsfxsize = 13cm
\epsffile{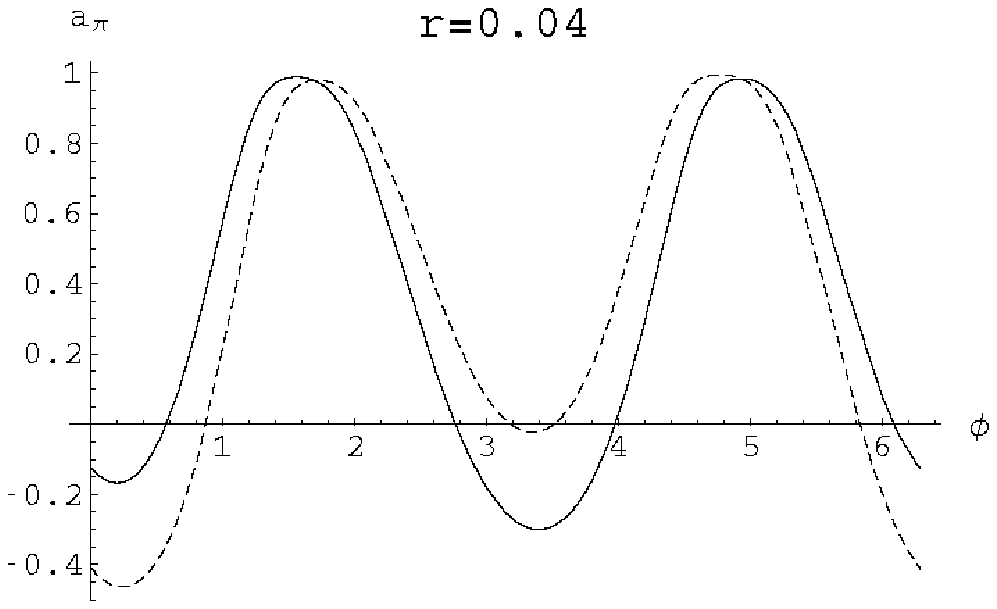}
\leavevmode
\end{center}
\caption{$a_\pi$ as a function of $\phi$ for $r=0.04$.
The dashed line  corresponds to the prediction obtained with  
the effective  Lagrangian previously used 
(without the linear term), while the
solid line corresponds to our new result.}
\end{figure}

\pagebreak

\begin{figure}
\begin{center}
\epsfxsize = 13cm
\epsffile{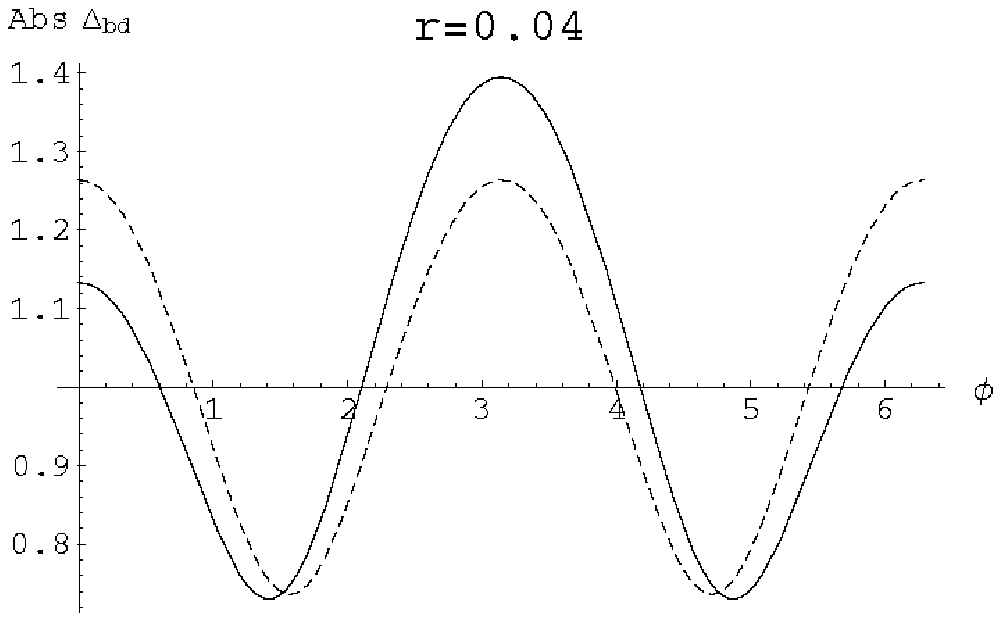}
\leavevmode
\end{center}
\caption{$\mid \Delta_{bd}\mid$ as a function of $\phi$  for $r=0.04$.
The dashed line  corresponds to the prediction obtained with  
the effective  Lagrangian previously used 
(without the linear term), while the
solid line corresponds to our new result.}
\end{figure}

\clearpage

\begin{figure}
\begin{center}
\epsfxsize = 13cm
\epsffile{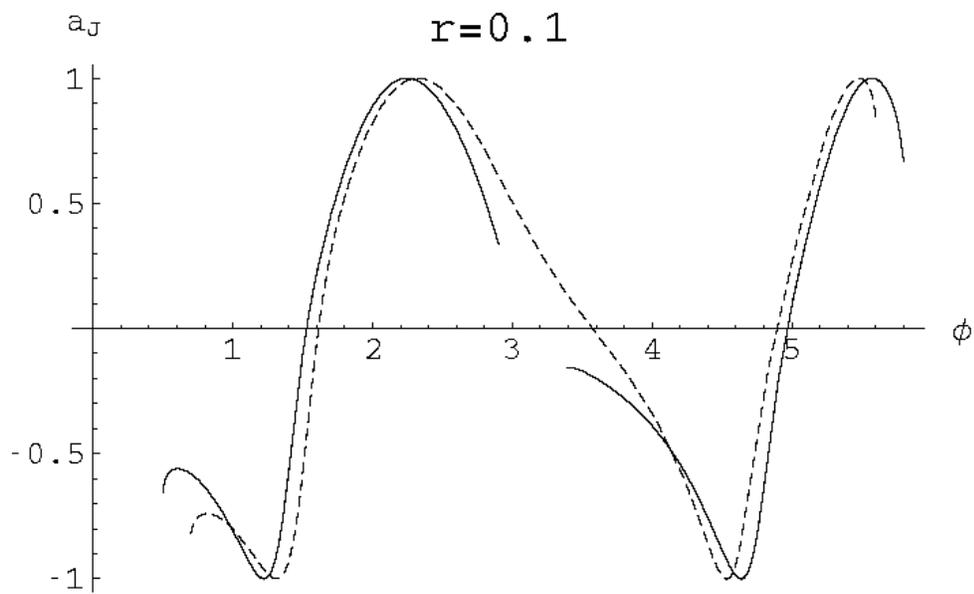}
\leavevmode
\end{center}
\caption{The same as figure 5 but with  $r=0.1$. It is important to 
notice here that not all the values of $\phi$ are allowed. }
\end{figure}

\clearpage

\begin{figure}
\begin{center}
\epsfxsize = 13cm
\epsffile{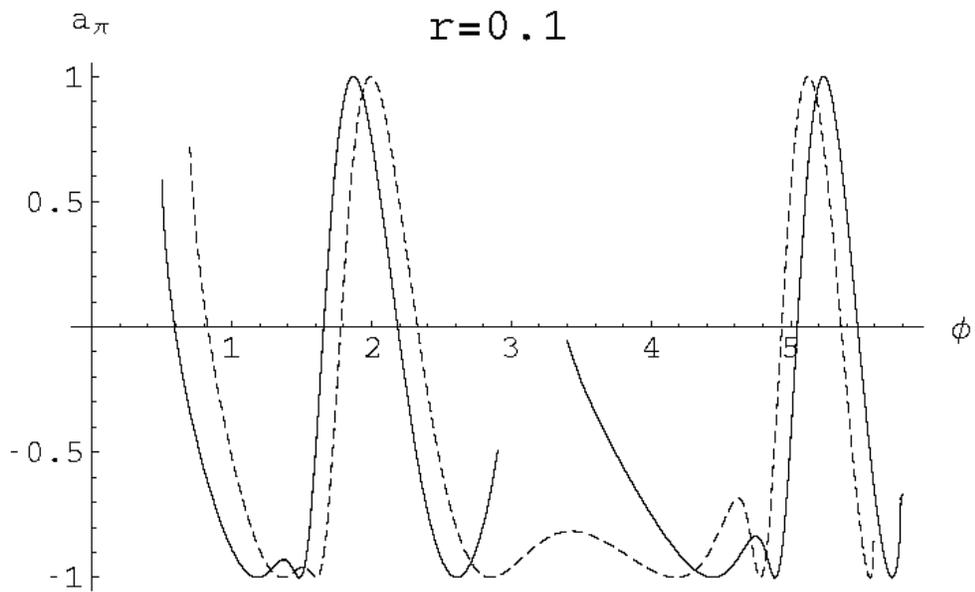}
\leavevmode
\end{center}
\caption{The same as figure 6 but with $r=0.1$. Again some $\phi$
values are forbidden.}
\end{figure}

\end{document}